\documentclass[aps,prl,twocolumn,showpacs,floatfix]{revtex4}
\usepackage{amsmath}
\usepackage{graphicx}
\usepackage{amssymb}
\begin{document}

\title{Nonequilibrium ``melting'' of a charge density wave insulator via an ultrafast laser pulse}
\author{Wen Shen$^1$}
\author{Yizhi Ge$^1$}
\author{A. Y. Liu$^1$}
\author{H. R. Krishnamurthy$^{2,3}$}
\author{T. P. Devereaux$^{4,5}$}
\author{J. K. Freericks$^1$}
\affiliation{
$^{1}$ Department of Physics, Georgetown University, Washington D.C., 20057, USA\\
$^2$Jawaharlal Nehru Centre for Advanced Scientific Research, Bangalore 560012, India\\
$^3$Department of Physics, India Institute of Science, Bangalore 560012, India\\
$^4$SLAC National Accelerator Laboratory, Stanford University, Stanford, CA 94305, USA\\
$^5$Geballe Laboratory for Advanced Materials, Stanford University, Stanford, CA 94305, USA
}
\begin{abstract}
We employ an exact solution of the simplest model for pump-probe time-resolved photoemission spectroscopy in charge-density-wave systems to show how, in nonequilibrium the gap in the density of states disappears while the charge density remains modulated, and then the gap reforms after the pulse has passed. This nonequilibrium scenario qualitatively describes the common short-time experimental features in TaS$_2$ and TbTe$_3$ indicating a quasiuniversality for nonequilibrium ``melting'' with qualitative features that can be easily understood within a simple picture.
\end{abstract}

\maketitle

The theory of second-order equilibrium phase transitions has a long history and is now well understood \cite{chaiken_lubensky}. In the case of electronic phase transitions induced by electron-phonon interactions, gaps in the electronic excitation spectrum typically open up simultaneously with the formation of long range charge density wave (CDW) or superconducting order in weak coupling, and precede the formation of these orders in strong coupling. Recent experiments in ultrafast pump/probe spectroscopy that investigated layered CDW materials, have revealed a new nonequilibrium paradigm where long-range CDW order persists but, by conventional interpretation, the local electronic excitation spectrum becomes gapless (by creating subgap states) for a transient period of time  \cite{demsar,brouet,perfetti1,perfetti3,schmitt1,eichberger,hellmann,schmitt2,petersen,dean,rohwer}. The similarity of these experiments to each other for quite different materials points towards a quasi-universal behavior in nonequilibrium, whose main features are captured with a simple exactly solvable model that we present below.

We concentrate on two materials that have been measured experimentally. The quasi two-dimensional material 1T-TaS$_2$ orders in a three sublattice star-of-David pattern, and develops an insulating gap that is believed to be due to strong electronic correlations \cite{fazekas,wu,kim}. In TbTe$_3$, the system condenses in unidirectional incommensurate CDW stripes, and the order only partially gaps the Fermi surface, leaving the system with metallic conductivity \cite{schmitt1,schmitt2,fang,brouet2}. In both materials, experiments have clearly shown the transient collapse of the CDW gap in the density of states (DOS) as the system is pumped into a nonequilibrium state by a large amplitude pulse \cite{perfetti1,perfetti3,schmitt1,schmitt2,petersen,dean}. At picosecond times after the pump pulse, the DOS, as inferred from the time-resolved photoemission spectroscopy (PES), oscillates due to coupling with the soft phonon that is involved in the CDW transition, before the gap fully reforms at longer times. In TaS$_2$, a pump/probe core-level x-ray photoemission spectroscopy experiment \cite{hellmann} further shows that the amplitude of the electronic CDW order parameter (given by the difference of the electric charge density on the different sublattices of the CDW) is reduced by the pulse, but does not vanish; it eventually settles into a long-time value that is reduced from the original size due to heating of the system and relaxation back to equilibrium. {\it Hence, unlike in equilibrium, experiments show that the electronic CDW order parameter and the gap in the electronic energy spectrum become partially decoupled in the sense that the gap collapses while residual modulated CDW order remains}.

This change in character of the many-body state has been named a nonequilibrium melting (or phase transition) of the CDW but it has an inherently different character from the equilibrium phase transitions. For example, the gap is initially fragile to the presence of an electric field while the electronic order parameter is initially robust (due to the frozen in lattice distortion yielding an inhomogeneous potential for the electrons), so the behavior of the system during the transient closing of the gap lies in a different physical realm than does the equilibrium phase transition.

Electrons interact on timescales on the order of a femtosecond ($h/1$~eV$= 4.2$~fs). This motivates the notion of a hot-electron model \cite{allen,perfetti1,perfetti3,jkf_trpes1}, where the electrons rapidly thermalize amongst themselves forming a hot quasi-thermal gas that equilibrates with the phonons on longer (typically picosecond) time scales \cite{eichberger,hertel}. The long-time behavior of our system resembles this hot-electron model  (even though we have no scattering that gives rise to thermalisation), but the most interesting regime is the short-time transient one where properties change most rapidly in time and are not described by such simplifications.

\begin{figure}[!ht]
\begin{center}
\includegraphics[scale=0.35]{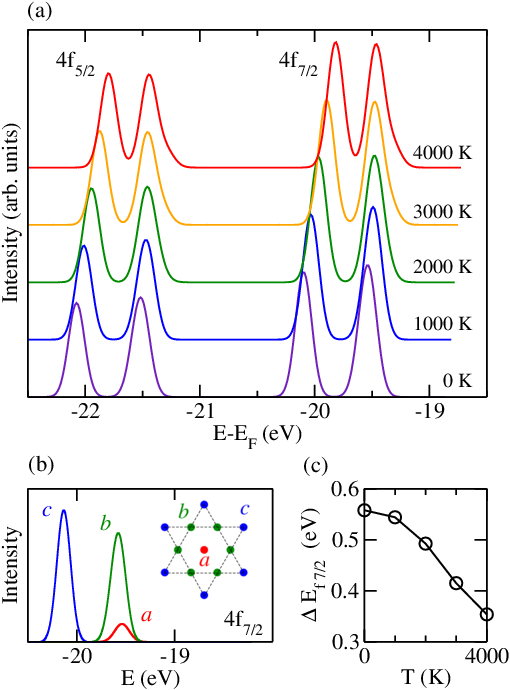}
\end{center}
\caption{(Color online) Calculated Ta $4f$ core-level photoemission spectrum for the commensurate CDW phase of 1T-TaS$_2$. (a) Dependence of the spectrum on electronic temperature in the hot electron model. The $f_{5/2}$ and $f_{7/2}$ levels are split due to spin-orbit coupling (2 eV) and  further split due to the existence of three inequivalent Ta sites in the CDW phase (0.5 eV). (b) Site-projected contributions to the $f_{7/2}$ spectrum. The 4f levels on Ta a and b sites lie close in energy and combine into a single peak. Inset is the star of David structure for the CDW (a, red; b, green; and c, blue). (c) Electronic-temperature dependence of the splitting between the a/b and c peaks in the $4f_{7/2}$ spectrum.
\label{fig: core}}
\end{figure}

Before introducing the simplified model for the CDW, we discuss how the electronic order parameter can survive to hot electron temperatures with a real materials calculation for TaS$_2$.
In a CDW system pumped by a laser pulse, the lattice distortion is frozen in at short times and cannot relax \cite{eichberger}. Hence the electrons always see a corrugated potential due to the ordered arrangement of the ion cores \cite{mazin} and respond to that potential by modulating their charge distribution so that the electronic order parameter does not vanish. We illustrate this in Fig.~\ref{fig: core}, where we calculate the core-level x-ray photoemission spectroscopy for TaS$_2$ with the lattice distortion fixed at its $T = 0$ value, but the electrons raised to temperatures up to 4000 K. The core-level shift is proportional to the CDW order parameter of the conduction electrons, indicating that the order decreases but remains for any finite temperature, as seen in experiment \cite{hellmann}.  These results were calculated within density-functional theory (DFT) using the all-electron, full potential code \cite{schwarz} WIEN2K. The Perdew-Burke-Ernzerhof \cite{perdew} version of the generalized gradient approximation was used to take into account the electron-electron interaction at the standard level done in DFT (additional details are described in~\cite{supplemental}).

\begin{figure}[!ht]
\includegraphics[scale=0.31]{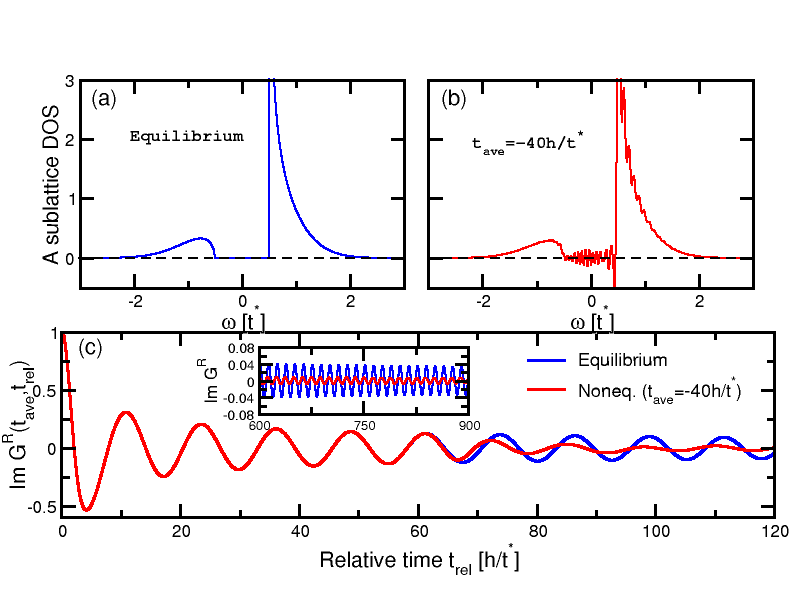}
\caption{(Color online) (a) The equilibrium local DOS on the A sublattice. (b) The nonequilibrium local DOS for $E_0=0.75$ at an average time $-40 h/t^*$.  (c) Imaginary part of the retarded Green's function in real time. The inset shows that the retarded Green's function has a long temporal tail.  
\label{fig: retarded}}
\end{figure}

 In nonequilibrium, one employs the Peierls substitution, which is a time-dependent shift of the momentum in the Hamiltonian.  Hence,  the instantaneous spectra is independent of the field and always has a full gap. But the nonequilibrium  DOS is determined by the retarded Green's function which is influenced by how the eigenfunctions change as a function of time, and often has its gap region modified. 
We examine the simplest model of a CDW insulator, where the ordering is driven by a periodic potential which is equal to zero on the B sublattice and equal to $U$ on the A sublattice of a bipartite lattice with equal numbers of A and B sites (the electrons are at half-filling to form an insulator)~\cite{supplemental}. The Hamiltonian, using standard notation for the creation and annihilation operators is
\begin{equation}
\mathcal{H}(t)=-\sum_{ij}\tau_{ij}(t) c^\dagger_i c^{}_j +(U-\mu)\sum_{i\in A} c^\dagger_ic^{}_i
-\mu\sum_{i\in B}c_i^\dagger c^{}_i
\end{equation}
with the chemical potential satisfying $\mu=U/2$ for half-filling and the time-dependent hopping in the presence of the pump pulse is given by the Peierls substitution as described in the supplemental material~\cite{supplemental}.
The fixed underlying potential in our model mimics the lattice distortion which does not change for short transient times. In equilibrium, the DOS develops a gap of magnitude equal to $U$ (we choose $U =1t^*$ throughout here with $t^*$ the renormalized hopping for a hypercubic lattice in infinite dimensions and the initial temperature before the pulse is equal to zero). The DOS is reflection symmetric about zero energy on the A and the B sublattices [see Fig.~\ref{fig: retarded}(a)]. On one lattice, there is a pile-up of states which gives a divergence that grows like the inverse square root at the lower gap edge, while on the other, the singularity lies at the upper gap edge. The Fourier transform of such a DOS can be shown to oscillate with an amplitude that decays with an inverse square root of time.

We model the laser pulse by a one and a half cycle electric field that approximately runs from a time of $-15h/t^*$ to $15h/t^*$ with an amplitude $E_0$ at $t=0$ that can be adjusted and is shown in Fig.~\ref{fig: summary}(d).
The exact solution of this nonequilibrium CDW (including a solution for the pump-probe PES signal \cite{jkf_trpes2}) requires one to use a Kadanoff-Baym-Keldysh formalism \cite{kadanoff_baym,keldysh,mahan} on a contour that runs from $t = -\infty$ to large positive time, returning back to $t = -\infty$. We solve for the retarded and lesser Green's functions which depend on two times \cite{kadanoff_baym} ($t$ and $t'$) using an exact evolution operator that can be expressed as a direct product of $2 \times 2$ matrices for each momentum point in the small Brillouin zone, and evaluated via the Trotter formula \cite{supplemental}. The solution is complicated because the different terms in the Trotter formula do not commute. 

We redefine the Green's functions in terms of the Wigner coordinates corresponding to the average $t_{ave} = (t + t')/2$ and relative $t_{rel} = t - t'$ times. The transient DOS at $t_{ave}$ is then defined to be the imaginary part of the Fourier transform of the retarded Green's function with respect to $t_{rel}$. Similarly, the time-resolved PES response function is found from a probe-pulse-weighted Fourier transform of the lesser Green's function \cite{jkf_trpes1,jkf_trpes2}. The Green's functions are nonlocal in time and they have long tails as functions of relative time, hence the effect of the pump pulse can modify them both at the time of the pulse, and for long average times before or after the pulse has ended. 

We illustrate this with the DOS in Fig.~\ref{fig: retarded}, which shows how the gap region can be significantly changed {\it even for average times far before the pulse is turned on}. Note that because this is a nonequilibrium situation, the DOS can become negative (due to no Lehmann representation for transient times), implying that the standard interpretation of the DOS fails for transient times. For example, even at a time $40h/t^*$ before the center of the pulse acts on the system ($25h/t^*$ before the pulse starts), we can see that the gap region of the DOS is significantly changed [Fig.~\ref{fig: retarded}(b)]. One can directly trace this to the change of the Green's function starting at a relative time of about $60h/t^*$ and continuing to long times [we fit the tail to $\omega\approx 10^6h/t^*$ to get a high-quality Fourier transform; the wiggles in the data in Fig.~\ref{fig: retarded}(c) are a real effect and not the result of any truncation]. 
The Green's function ``feels'' the effect of the field, because one time is before and one after the field has been turned on, leading to the fragile gap in the DOS.

\begin{figure}[htb]
\includegraphics[scale=0.28]{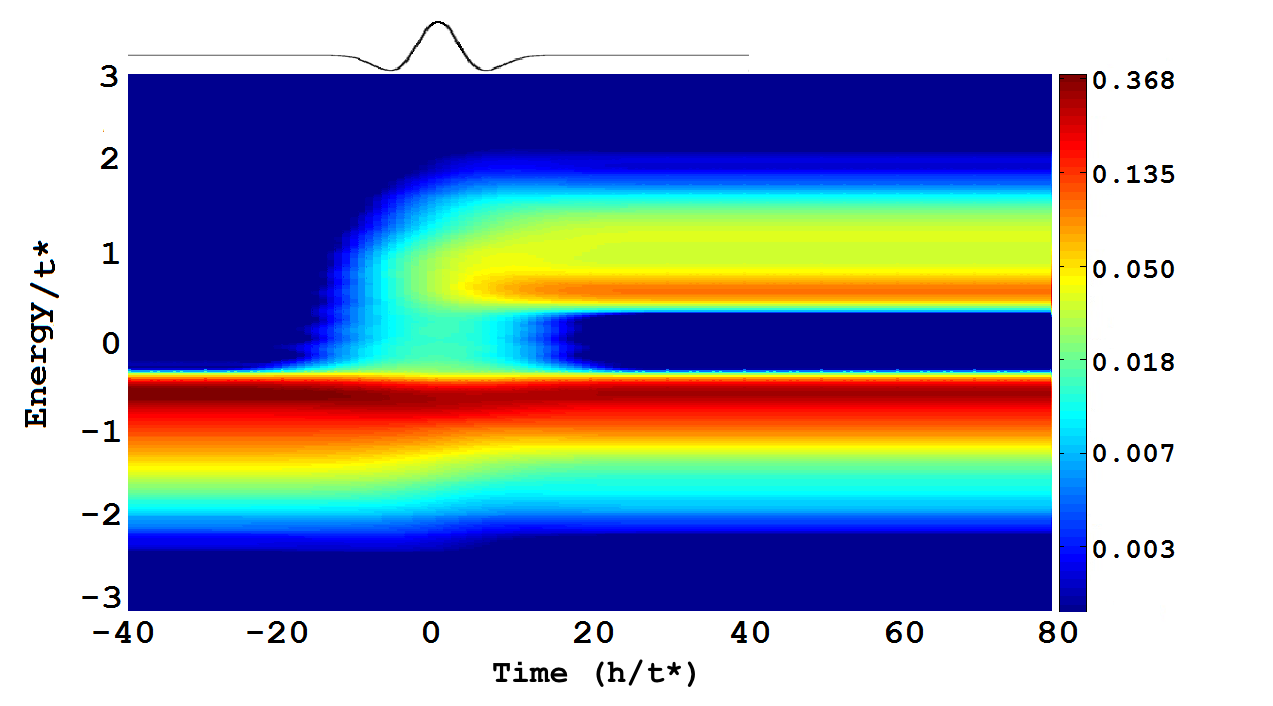}
\caption{(Color online) Calculated time-resolved PES at $T = 0$ with $E_0=0.75$ and averaged over the $A$ and $B$ sublattices, plotted in false color. The electric field is shown above the plot.  Movies of this time evolution can be found with the supplemental material for $E_0=0.75$, 1.0, and 1.25~\cite{supplemental}.
\label{fig: tr_pes}}
\end{figure}

\begin{figure}[!ht]
\includegraphics[scale=0.36,clip]{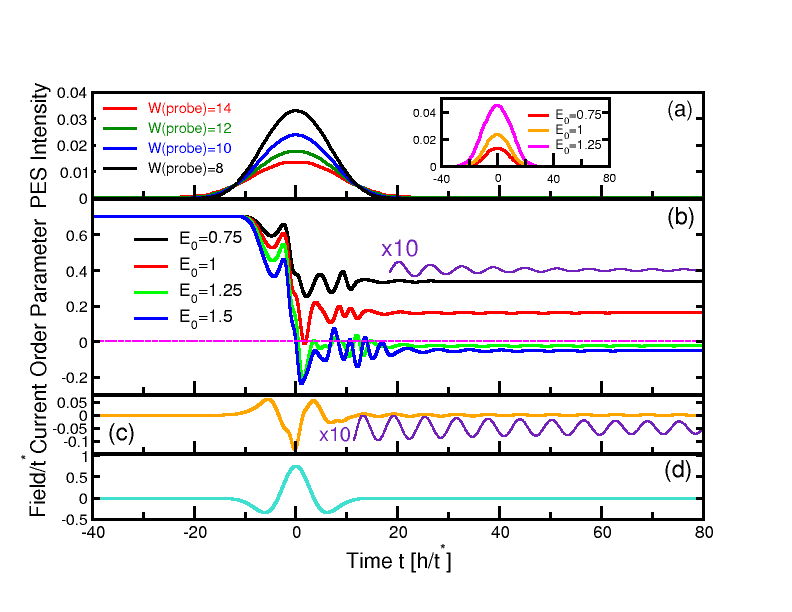}
\caption{(Color online) (a) Calculated time-resolved PES signal at the Fermi energy with different probe pulse widths and $E_0=0.75$.  Inset: the same result for the widest width probe pulse but different $E_0$ values. (b) Conduction electron order parameter for the CDW as a function of time for different pulse amplitudes (zero is indicated by the magenta dashed line). (c) Transient current for $E_0=0.75$.  With a small field, the current follows in phase with the electric field. For time delays after the pump pulse, the oscillations in the current have the same frequency as the order parameter has ($2\pi/U$, enhanced purple dashed line). (d) Pump pulse electric field with amplitude $E_0=0.75$. 
\label{fig: summary} }
\end{figure}

We show the time-resolved PES signal for an electric pulse amplitude satisfying $E_0=0.75$ in Fig.~\ref{fig: tr_pes}, using a probe pulse that has a Gaussian envelope with a width of $14h/t^*$ (the ``gap'' in the time-resolved PES signal is more robust than in the DOS due to the finite width of the probe pulse which limits the range in time for the Fourier transform, bypassing some effects due to the long tails, and the fact that the PES signal is always manifestly nonnegative).  One can see that in the range of average time from about $-20h/t^*$ up to about $20h/t^*$ the gap disappears, and then reforms for longer times. We also see a significant transfer of electrons from the lower to the upper band due to the nonequilibrium pumping of energy into the system. The system does tend toward a steady state at long times, but it isn't thermal because there is no electron scattering to thermalize the excited electrons. These results are similar for $E_0=1$ and $E_0=1.25$~\cite{supplemental}.

\begin{figure}[!ht]
\begin{center}
\includegraphics[scale=0.36,clip]{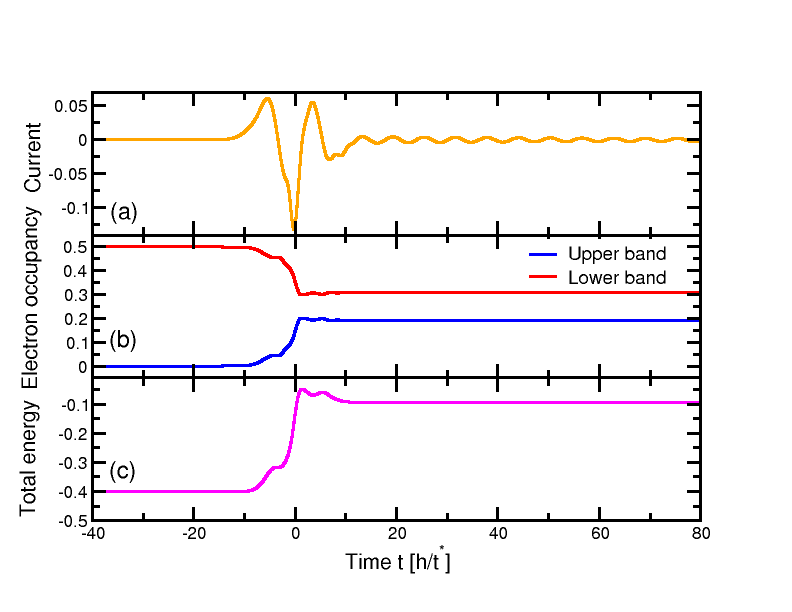}
\end{center}
\caption{(Color online) Calculated current (a), transient filling of the upper (red) and lower (blue) bands (b) and total energy (c) for the CDW system with U=1 and $E_0=0.75$.
\label{fig: energy}}
\end{figure}

Figures~\ref{fig: summary} and \ref{fig: energy} provide summaries of the various features of the nonequilibrium ``phase transition''. Panel (a) shows the PES signal at $\omega = 0$ as a function of time for a range of different probe pulse widths (full width at half max) at a pulse amplitude $E_0=0.75$. One can clearly see that the suppression of the gap (by filling in subgap states) is robust once the width is large enough. The inset shows how the gap magnitude grows with the field amplitude, but the width in time remains the same. Panel (b) shows that the electronic CDW order parameter of the conduction electrons  is reduced due to the pump pulse but does not vanish at $E_0=0.75$. For $E_0=1$, the order parameter barely touches zero for an instant and then recovers to a small positive value. For $E_0=1.25$ and $1.5$, a small {\it reversal of the order} is found in this system which remains reversed after the electric field pulse has gone. The steady state arises because this system is a closed system and the only exchange of energy occurs when the electric field pulse is present. All of these order parameters have oscillations with a period of $2\pi /U$ at long times, which eventually are dephased, as does the current in panel (c) (the definition is given in the supplementary material~\cite{supplemental}).

In  Fig.~\ref{fig: energy}, we see complex behavior for the expectation value of the total energy $\mathcal{E}(t)=\langle\mathcal{H}(t)\rangle$ while the pulse is on ($E_0=0.75$), eventually settling down to a constant value. We can use this steady-state energy to estimate the effective temperature of the system, by calculating the energy as a function of temperature for the system in equilibrium, and setting the temperature by equating to the  nonequilibrium energy after the pulse (yields $k_BT=2.38t^*$). We also can calculate the filling in the lower and the upper bands of the CDW as a function of time for the nonequilibrium case [panel (b) of Fig.~\ref{fig: energy}], and compare it to the equilibrium fillings in each band as a function of temperature. This yields $k_BT=1.67t^*$ (for $E_0=1$ the energy temperature is $3.91t^*$ and the filling one is $3.33t^*$ while for $E_0=1.25$ the energy temperature is $31.5t^*$ and the filling one is $-2250t^*$). In thermal equilibrium, these two temperatures must agree. Their difference is one measure of the nonthermal nature of the final state. What is remarkable is that this noninteracting system can remain fairly close to a thermal distribution (but this only holds for small amplitudes of the field).  The energy approaches a constant before the current does, because no energy can be added to the system when the electric field vanishes since $d\langle \mathcal{H}(t)\rangle/dt=-{\bf E}(t)\cdot\langle{\bf j}(t)\rangle$. 

\paragraph{Conclusion}
In this work we have shown that a new paradigm exists in nonequilibrium, where driving a CDW system with large fields can cause the gap to transiently vanish for intermediate times in the presence of a generically reduced order parameter. We find this behavior with an exact solution of a simplified model of a CDW insulator described by a staggered potential that is different on one of two sublattices. While this exact solution does not capture all of the quantitative details of the experiments (particularly when coupling of electrons to phonons becomes important), for short and intermediate times, it provides a consistent qualitative explanation of the experimental data and is consistent with the emergence of a quasi-universal behavior in nonequilibrium. The timescale for the reopening of the gap is smaller in this model than in experiment because we have not included a phonon bath that couples to the electrons via the electron-phonon coupling. This allows for an oscillating transfer of energy back and forth between electrons and phonons, until they are damped. When the phonons transfer energy back to the electrons, it is similar to repumping the electrons by an external pulse, which we believe is why the gap remains closed for longer time in experiments.
\paragraph{Acknowledgments}
The calculations of the core-level x-ray PES in the hot electron model was supported by the National Science Foundation under Grant No. DMR-1006605. The development of the parallel computer algorithms for the time-resolved PES calculations was supported by the National Science Foundation under Grant No. OCI-0904597. The data analysis and application to experiment was supported by the Department of Energy, Office of Basic Energy Research under Grants No. DE-FG02-08ER46542 (Georgetown), DE-AC02-76SF00515 (Stanford/SLAC), and DE-FG02-08ER46540 and DE-SC0007091 (for the collaboration). High performance computer resources utilized the National Energy Research Scientific Computing Center supported by the Department of Energy, Office of Science, under Contract No. DE- AC02-05CH11231. J.K.F. was also supported by the McDevitt bequest at Georgetown. H.R.K. acknowledges support from DST (India). The Indo-US collaboration was supported by the Indo-US Science and Technology Forum under a center grant numbered JC-18-2009.

\newpage

\section{Supplemental information}

\paragraph*{Technical details for the density functional theory calculation}
 The Ta $4f$ states were treated as part of the semi-core, and a second variational approach built on the scalar relativistic approximation was used to include spin-orbit coupling effects on the valence and semi-core states. For this hot-electron model, the $\sqrt{13}\times\sqrt{13}\times 1$ CDW structure was first fully relaxed using a small electronic temperature. Then the structural parameters were kept fixed while the ``hot'' electronic temperature was raised. Fig.~1 of the article shows the density of states, which approximates the core-level photoemission spectrum under the assumption that matrix elements are nearly constant (the calculated spectra were broadened using a Gaussian width of 0.068~eV).

\paragraph*{Formalism for the exact nonequilibrium solution in the presence of CDW order.}
The retarded and lesser Green's functions are defined as a quantum statistical average of time dependent creation and annihilation operators in the Heisenberg representation by the following formulas:
 \begin{equation}
G^R_{ij} (t,t')=-i \theta (t-t')\langle\{c_i^{}(t) , c_j^\dagger (t')\}_+\rangle 
\end{equation}
\begin{equation}
G^{<}_{ij}(t,t')=i\langle c_j^{\dagger}(t')c_i^{}(t)\rangle .
\end{equation}
Here $c_i^{}(t)$ and $c^{\dagger}_i(t)$ are creation and annihilation operators for a spinless fermion at lattice site $i$ (in the Heisenberg representation) and the curly brackets denote an anticommutator. The angle brackets denote a trace over all quantum states weighted by the equilibrium density matrix that the system is initialized in in the distant past ($t\rightarrow -\infty$ and $t'\rightarrow -\infty$) and is given by $\langle\cdot\rangle={\rm Tr}\rho\cdot$ with $\rho=\exp[-\beta\mathcal{H}(t\rightarrow -\infty)]/\mathcal{Z}$ and $\mathcal{Z}={\rm Tr} \exp[-\beta\mathcal{H}(t\rightarrow -\infty)]$. We also use similar formulas in momentum space, where the change in labels from real space to momentum space will be obvious. Note that the retarded Green's function is nonzero only for positive relative time $t-t'>0$, while the lesser Green's function has no such restriction. The Green's functions are employed to calculate the DOS, the electron concentration on the $A$ and $B$ sublattices (order parameter), the current, the total energy, and the time-resolved PES signal.

Our model is that of spinless electrons moving on a lattice with a bipartite structure (like a checkerboard) that has a different site energy on the two sublattices. The time-dependent Hamiltonian in the Schr\"odinger picture is
 \begin{eqnarray}
\mathcal{H}(t)&=&-\displaystyle\sum_{\langle ij\rangle}\left [\tau_{ij}(t)\emph{c}_{i}^{\dagger}\emph{c}_{j}^{}+ h.c.\right ]+\displaystyle\sum_{i\in{A}}( U-\mu)\emph{c}_{i}^{\dagger}\emph{c}_{i}^{}\nonumber\\
&+&\displaystyle\sum_{i\in{B}}(-\mu)\emph{c}_{i}^{\dagger}\emph{c}_{i}^{}.
\end{eqnarray}
The first term is the kinetic energy, which involves a hopping between nearest neighbor lattice sites $i$ and $j$ with a time-dependent hopping integral $\tau_{ij}(t)$ (the symbol $\langle ij\rangle$ indicates a summation over nearest neighbor pairs and h.c.~is the Hermitian conjugate). The second and third terms include the chemical potential $\mu$ and the external potential $U$ which is nonzero only on the $A$ sublattice. We will set $\mu=U/2$ in our calculations to have the case of half filling. In equilibrium, the hopping integral is a constant, but in nonequilibrium, it becomes time-dependent due to the Peierls' substitution which describes the pump pulse. The time-dependent hopping integral is then
\begin{eqnarray}
\tau_{ij}(t)&=& \frac{t^*}{2\sqrt{d}} \exp\left[-\frac{ie}{\hbar c}\int_{{\bf R}_i}^{{\bf R}_j}\textbf {A} (\emph{t})\cdot d\textbf {r}\right]\nonumber\\
&=& \frac{t^*}{2\sqrt{d}}\exp\left[\frac{ie}{\hbar c}\textbf {A} (\emph{t})\cdot ({\bf R}_i -{\bf R}_j)\right]
\end{eqnarray}
and we will be taking the limit as $d\rightarrow\infty$ (this is done solely for convenience, as all of the formalism holds in arbitrary dimensions).
Here ${\textbf {A} (\emph{t}})$ is the time dependent (but spatially uniform) vector potential in the Hamiltonian gauge (where the scalar potential vanishes). The pump pulse is taken in the diagonal direction ${\textbf {A} (\emph{t}})=A(t)(1,1,1\dots)$ with $A(t)=-E_0t\exp(-t^2/25)$, and $E_0$ being the magnitude of the field at $t=0$, which corresponds to the maximum amplitude of the field
[${\bf E}(t)=-\partial {\bf A}(t)/c\partial t$]. The symbol ${\bf R}_i$ denotes the position vector of the $i$th lattice site. We work in units where $e=\hbar =c=1$.

In this case, the (time-dependent) electronic band structure becomes 
 \begin{equation}
\varepsilon\left(\textbf{k}-\frac{e\textbf{A} (t)}{\hbar c}\right)=-\lim_{d\rightarrow\infty}\frac{t^*}{\sqrt{d}}\displaystyle\sum_{l=1}^{d}\cos\left[a\left(k_l-A_l(t)\right)\right]
\end{equation}
which we will denote as $\epsilon({\bf k};t)$.
With the field pointing along the diagonal, the time-dependent bandstructure can be represented in terms of two effective band energies: (i) $\epsilon({\bf k})=-\lim_{d\rightarrow\infty}t^*\sum_{l=i}^d\cos(k_la)/\sqrt{d}$, which is the ordinary bandstructure and (ii) $\bar\epsilon({\bf k})=-\lim_{d\rightarrow\infty}t^*\sum_{l=i}^d\sin(k_la)/\sqrt{d}$, which is the projection of the equilibrium electron velocity along the direction of the field. We will work in units where the lattice constant satisfies $a=1$.

In equilibrium, the Green's functions are straightforward to solve because the system consists of noninteracting electrons on a lattice with a basis. In nonequilibrium, one needs to properly include the time dependence, which is easiest to do with {\it a time-independent basis} for the creation and annihilation operators (otherwise one needs to take into account the derivative of how the basis changes with time).   In momentum space, the time-evolution operator $U(t,t';{\bf k})$ is a $2\times 2$ matrix that satisfies the following equations:
\begin{equation}
i\partial{c_{\bf k}(t)}/\partial{t}=-\left[H(t),c_{\bf k}(t)\right]  ,
\end {equation}
\begin{equation}
i\partial{c_{{\bf k}+{\bf Q}}(t)}/\partial{t}=-\left[H(t),c_{{\bf k}+{\bf Q}}(t)\right]  ,
\end{equation}
\begin{equation}
\left(\begin{array}{c}c_{\bf k}(t)\\c_{{\bf k}+{\bf Q}}(t)\end{array}\right)=U(t,t';{\bf k})\left(\begin{array}{c}c_{\bf k}(t')\\c_{{\bf k}+{\bf Q}}(t')\end{array}\right),
\end{equation}
with ${\bf Q}=(\pi,\pi,\pi,\ldots)$ the ordering wavevector for the CDW.
The evolution operator can be constructed via the Trotter formula for $2\times 2$ matrices
\begin{eqnarray}
U(t,t';{\bf k})&=&U(t,t-\Delta t;{\bf k})U(t-\Delta t,t-2\Delta t;{\bf k})\cdots\nonumber\\
&\times& U(t'+\Delta t, t';{\bf k}).
\end{eqnarray}
The explicit formulas simplify at half filling (where $\mu=U/2$), resulting in
\begin{eqnarray}
 &~&U(t,t-\Delta t; {\bf k})=\cos \left ( \Delta t\sqrt{\epsilon^2({\bf k};\bar t)+\frac{U^2}{4}}\right )\mathbb{I}\\
&-&i\left ( 
\begin{array}{c c}
          \epsilon({\bf k};\bar t)&\frac{U}{2}\\
\frac{u}{2}&-\epsilon({\bf k};\bar t)
         \end{array}
\right )\frac{\sin\left (\Delta t\sqrt{\epsilon^2({\bf k};\bar t)+\frac{U^2}{4}}\right )}{\sqrt{\epsilon^2({\bf k;\bar t})+\frac{U^2}{4}}},\nonumber
\end{eqnarray}
and $\bar t=t-\Delta t/2$, with $\Delta t$ the discretization time step for real times along the Keldysh contour. 
Note that because we use an exact expression for the exponential of the $2\times 2$ matrix that appears in the time-ordered product, the evolution operator is always unitary (even for large $\Delta t$), although the accuracy is still determined by the size of $\Delta t$.

The continuous time-ordered product follows in the limit as the time step $\Delta t$ goes to zero but we evaluate with a fixed time constant when we perform the calculations numerically.
Using this result, the retarded Green's function becomes
\begin{eqnarray}
G^R_{ii} (t,t')&=&-i \theta (t-t')\displaystyle\sum_{\bf k}\{U_{11}(t,t';{\bf k})+U_{22}(t,t';{\bf k})\nonumber\\
&\pm& U_{12}(t,t';{\bf k})\pm U_{21}(t,t';{\bf k})\}
\end{eqnarray}
where the subscripts are for the $2\times 2$ matrix and the $\pm$ sign is for the A (B) sublattice, respectively. In infinite dimensions, with the electric field oriented along the diagonal, the sum over momentum can be replaced by a two-dimensional gaussian-weighted integral over the two band energies $\epsilon$ and $\bar\epsilon$.  Note that the retarded Green's function depends only on the relative time between $t$ and $t'$ and hence has no memory of the history of the evolution of the system. As a check of our formulas, we verify the nonequilibrium moment sum rules for the retarded Green's functions and find that they are exactly satisfied for the zeroth and first three positive power moments~[S1].

Once $G_{ii}^R(t_{ave},t_{rel})$ is known in terms of the Wigner coordinates of average and relative time, the local DOS [$A_{A,B}(w)=-\frac{1}{\pi}G^R_{AA,BB} (\omega)$] is found by Fourier transformation with respect to relative time via
\begin{equation}
G^R_{ii} (t_{ave},\omega)=\int dt_{rel} e^{i\omega t_{rel}}G^R_{ii}(t_{ave},t_{rel}).
\end{equation}
We can only directly calculate the retarded Green's function for moderate times (on the order of 1200 $h/t^*$) which is often insufficient to determine the Fourier transform due to the slow decay of the Green's function. We find, however, that we can fit the Green's function over a wide range of times by the form
\begin{equation}
{\rm Im}G^R_{ii}(t_{ave},t_{rel})=\frac{a_0}{\sqrt{t_{rel}}}\cos(0.25t_{rel}+a_1)+a_2
\end{equation}
with a similar result for the real part. We extrapolate the fits out to
$t_{rel}\approx 10^6 h/t^*$ and then perform the Fourier transform for the results in Fig.~2 of the main text.

The lesser Green's function has a more complicated form because it does depend on the history of the evolution of the system and hence does not depend solely on evolution operators between times $t$ and $t'$, but rather depends on the evolution operators from a distant time in the past up to time $t$ and time $t'$. The formula for it is not too enlightening and is cumbersome, so we show it only for the $11$ or ${\bf kk}$ component (other cases follow similarly from a simple substitution of the evolution operators into the definition of the lesser Green's function):
\begin{eqnarray}
&~&G_{11}^<({\bf k},t,t')=\\
&~&U_{11}^\dagger(t_0,t';{\bf k})U_{11}(t,t_0;{\bf k})\langle c_{\bf k}^\dagger(t_0)c_{\bf k}(t_0)\rangle\nonumber\\
&+&
U_{11}^\dagger(t_0,t';{\bf k})U_{12}(t,t_0;{\bf k})\langle c_{\bf k}^\dagger(t_0)c_{{\bf k}+{\bf Q}}(t_0)\rangle\nonumber\\
&+&U_{21}^\dagger(t_0,t';{\bf k})U_{11}(t,t_0;{\bf k})\langle c_{{\bf k}+{\bf Q}}^\dagger(t_0)c_{\bf k}(t_0)\rangle\nonumber\\
&+&U_{21}^\dagger(t_0,t';{\bf k})U_{12}(t,t_0;{\bf k})\langle c_{{\bf k}+{\bf Q}}^\dagger(t_0)c_{{\bf k}+{\bf Q}}(t_0)\rangle , \nonumber
\end{eqnarray}
where $t_0\rightarrow-\infty$ is the initial time, and the four different expectation values are taken in the initial equilibrium state before the field is turned on.

The time-resolved PES response function is calculated as  a probe pulse weighted relative time Fourier transform of the lesser Green's function centered at time $t_p$,
\begin{equation}
P_{i}(\omega,t_p)=-i\int  dt \int  dt' s(t)s(t')e^{-i\omega (t-t')} G^<_{ii}(t+t_p,t'+t_p),
\end{equation}
where the response must be calculated for each sublattice (the experimental response will be the average over both sublattices).
Since the probe pulse provides a natural cutoff, no extrapolation to large times is needed for this calculation.
The probe pulse is approximated by a gaussian,
\begin{equation}
s(t)=\frac{1}{\sigma \sqrt\pi}\exp^{-t^2/\sigma^2}
\end{equation}
with width $\sigma$.
The narrower the probe pulse width, the better the time resolution and the worse the energy resolution, and {\it vice versa} for broader probe pulses.  

We also calculate the current, the total energy, and the occupancy in the upper/lower bands of the instantaneous band structure.  The formulas for these quantities can be expressed as equal time expectation values, which can be found in a straightforward fashion from the equal time lesser Green's function.  We write the formulas here in terms of the equal time expectation values.  The current satisfies
\begin{equation}
j(t)=\sum_{{\bf k}:\epsilon({\bf k})<0}\nabla \epsilon({\bf k};t)\left \langle \left (
c_{\bf k}^\dagger(t)c_{\bf k}(t)-c_{{\bf k}+{\bf Q}}^\dagger(t)c_{{\bf k}+{\bf Q}}(t)\right ) \right \rangle .
\end{equation}
The total energy becomes
\begin{eqnarray}
\mathcal{E}(t)&=&\sum_{{\bf k}: \epsilon({\bf k})<0}\left [
\epsilon({\bf k};t)\left (\left \langle c_{\bf k}^\dagger(t)c_{\bf k}(t)\right \rangle -
\left \langle c_{{\bf k}+{\bf Q}}^\dagger(t)c_{{\bf k}+{\bf Q}}(t)\right \rangle\right )\right .\nonumber\\
&+&\frac{U}{2}\left . \left (\left \langle c_{{\bf k}+{\bf Q}}^\dagger(t)c_{\bf k}(t)\right \rangle +
\left \langle c_{\bf k}^\dagger(t)c_{{\bf k}+{\bf Q}}(t)\right \rangle\right  )\right ] .
\end{eqnarray}
And the occupancy of the upper and lower instantaneous bands become
\begin{eqnarray}
n_+(t)&=&\sum_{{\bf k}: \epsilon({\bf k})<0}\left [
\alpha^2({\bf k};t)\left \langle c_{\bf k}^\dagger(t)c_{\bf k}(t)\right \rangle \right .\\
&+&
\beta^2({\bf k};t)\left \langle c_{{\bf k}+{\bf Q}}^\dagger(t)c_{{\bf k}+{\bf Q}}(t)\right \rangle\nonumber\\
&+&\alpha({\bf k};t)\beta({\bf k};t)\left . \left (\left \langle c_{{\bf k}+{\bf Q}}^\dagger(t)c_{\bf k}(t)\right \rangle +
\left \langle c_{\bf k}^\dagger(t)c_{{\bf k}+{\bf Q}}(t)\right \rangle\right  )\right ] \nonumber
\end{eqnarray}
and
\begin{eqnarray}
n_-(t)&=&\sum_{{\bf k}: \epsilon({\bf k})<0}\left [
\beta^2({\bf k};t)\left \langle c_{\bf k}^\dagger(t)c_{\bf k}(t)\right \rangle \right .\\
&+&
\alpha^2({\bf k};t)\left \langle c_{{\bf k}+{\bf Q}}^\dagger(t)c_{{\bf k}+{\bf Q}}(t)\right \rangle\nonumber\\
&-&\alpha({\bf k};t)\beta({\bf k};t)\left . \left (\left \langle c_{{\bf k}+{\bf Q}}^\dagger(t)c_{\bf k}(t)\right \rangle +
\left \langle c_{\bf k}^\dagger(t)c_{{\bf k}+{\bf Q}}(t)\right \rangle\right  )\right ] ,\nonumber
\end{eqnarray}
with the two time-dependent coefficients given by
\begin{equation}
\alpha({\bf k};t)=\frac{U/2}{\sqrt{2[\epsilon^2({\bf k};t)+U^2/4-\epsilon({\bf k};t)\sqrt{\epsilon^2({\bf k};t)+U^2/4}]}}
\end{equation}
and
\begin{equation}
\beta({\bf k};t)=\frac{-\epsilon({\bf k};t)+\sqrt{\epsilon^2({\bf k};t)+U^2/4}}{\sqrt{2[\epsilon^2({\bf k};t)+U^2/4-\epsilon({\bf k};t)\sqrt{\epsilon^2({\bf k};t)+U^2/4}]}}.
\end{equation}

\paragraph*{Computational algorithm and resources.}
Since each momentum point is independent of each other, we can easily parallelize the algorithm in the master/slave format, sending different momentum points to different CPUs. All of the times, however, need to be evaluated on each slave and stored, to be later accumulated when solving for local properties. We used the Cray XE6 (Hopper) at NERSC for the numerical work. This machine has 153,216 compute cores, 217 TB of memory and 2~PB of disk storage. Our code for the TR-PES is efficiently parallelized and typically ran on 10,000 processors; we used approximately 500,000 CPU-hrs for the project.

\paragraph*{Time-resolved PES movies.}
In Fig.~3 of the main text, we have already presented the time-resolved PES in a false-color plot for $E_0=1.25$. By taking vertical cuts through the data and converting from false color plots to line plots for each frame, we have constructed a movie of the time-resolved PES signal as a function of time (in steps of $1 h/t^*$). The file
{\it shensupplementalmovie\_E=XXX.gif}
shows the time-resolved PES with a probe width of 14 $h/t^*$. This animation starts  at a negative time of $-40$~$h/t^*$ and ends at a positive time of 40 $h/t^*$. Three movies are shown, with $E_0=0.75$, 1, and 1.25.

\subsection*{Supplemental Reference}
\begin{itemize}
\item[{[S1]}]
 V. M. Turkowski and J. K. Freericks, Phys. Rev. B {\bf 73}, 075108 (2006);  Erratum, Phys. Rev. B {\bf 73}, 209902(E) (2006);
V. M. Turkowski and J. K. Freericks,  Phys. Rev. B {\bf 77}, 205102 (2008); Erratum: Phys. Rev. B {\bf 82}, 119904(E) (2010);
J. K. Freericks and V. Turkowski,  Phys. Rev. B {\bf 80}, 115119 (2009); Erratum: Phys. Rev. B {\bf 82}, 129902(E) (2010).

\end{itemize}

\end{document}